\documentstyle[twoside,epsf]{article}
 
\catcode`\@=11
\long\def\@makefntext#1{ 
\protect\noindent \hbox to 3.2pt {\hskip-.9pt
$^{{\eightrm\@thefnmark}}$\hfil}#1\hfill} 
 
\def\thefootnote{\fnsymbol{footnote}}
 \def\@makefnmark{\hbox to 0pt{$^{\@thefnmark}$\hss}}  
 
\def\ps@myheadings{\let\@mkboth\@gobbletwo
\def\@oddhead{\hbox{} 
\rightmark\hfil\eightrm\thepage}
\def\@oddfoot{}\def\@evenhead{\eightrm\thepage\hfil 
\leftmark\hbox{}}\def\@evenfoot{}
\def\sectionmark##1{}\def\subsectionmark##1{}}
 
\oddsidemargin=\evensidemargin
\renewcommand{\thefootnote}{\fnsymbol{footnote}}

\newcounter{sectionc}\newcounter{subsectionc}\newcounter{subsubsectionc}
\renewcommand{\section}[1] {\vspace{12pt}\addtocounter{sectionc}{1}
\setcounter{subsectionc}{0}\setcounter{subsubsectionc}{0}\noindent
        {\tenbf\thesectionc. #1}\par\vspace{5pt}}
\renewcommand{\subsection}[1] {\vspace{12pt}\addtocounter{subsectionc}{1}
        \setcounter{subsubsectionc}{0}\noindent
        {\bf\thesectionc.\thesubsectionc. {\kern1pt \bfit #1}}\par\vspace{5pt}}
\renewcommand{\subsubsection}[1] {\vspace{12pt}\addtocounter{subsubsectionc}{1}
        \noindent{\tenrm\thesectionc.\thesubsectionc.\thesubsubsectionc.
        {\kern1pt \tenit #1}}\par\vspace{5pt}}

\newcounter{appendixc}
\newcounter{subappendixc}[appendixc]
\newcounter{subsubappendixc}[subappendixc]
\renewcommand{\thesubappendixc}{\Alph{appendixc}.\arabic{subappendixc}}
\renewcommand{\thesubsubappendixc}
        {\Alph{appendixc}.\arabic{subappendixc}.\arabic{subsubappendixc}}
 
\renewcommand{\appendix}[1] {\vspace{12pt}
        \refstepcounter{appendixc}
        \setcounter{figure}{0}
        \setcounter{table}{0}
        \setcounter{lemma}{0}
        \setcounter{theorem}{0}
        \setcounter{corollary}{0}
        \setcounter{definition}{0}
        \setcounter{equation}{0}
        \renewcommand{\thefigure}{\Alph{appendixc}.\arabic{figure}}
        \renewcommand{\thetable}{\Alph{appendixc}.\arabic{table}}
        \renewcommand{\theappendixc}{\Alph{appendixc}}
        \renewcommand{\thelemma}{\Alph{appendixc}.\arabic{lemma}}
        \renewcommand{\thetheorem}{\Alph{appendixc}.\arabic{theorem}}
        \renewcommand{\thedefinition}{\Alph{appendixc}.\arabic{definition}}
        \renewcommand{\thecorollary}{\Alph{appendixc}.\arabic{corollary}}
        \renewcommand{\theequation}{\Alph{appendixc}.\arabic{equation}}
        \noindent{\tenbf Appendix \theappendixc #1}\par\vspace{5pt}}
\newcommand{\subappendix}[1] {\vspace{12pt}
        \refstepcounter{subappendixc}
        \noindent{\bf Appendix \thesubappendixc. {\kern1pt \bfit #1}}
        \par\vspace{5pt}}
\newcommand{\subsubappendix}[1] {\vspace{12pt}
        \refstepcounter{subsubappendixc}
        \noindent{\rm Appendix \thesubsubappendixc. {\kern1pt \tenit #1}}
        \par\vspace{5pt}}
 
\newcommand{\textlineskip}{\baselineskip=13pt}
\newcommand{\smalllineskip}{\baselineskip=10pt}
 
\def\eightcopyright{\copyright}
 
\newcommand{\copyrightheading}[1]
        {\vspace*{-2.5cm}\smalllineskip{\flushleft
        {\eightrm International Journal of Modern Physics C, #1}\\
        {\eightrm $\eightcopyright$\, World Scientific Publishing
         Company}\\
         }}
 
\newcommand{\publisher}[2]{{\begin{center}\eightrm\smalllineskip
        Received #1\\
        Revised #2
        \end{center}
        }}
 
\def\abstracts#1#2#3{{
        \centering{\begin{minipage}{4.5in}\baselineskip=10pt\eightrm
        \parindent=0pt #1\par
        \parindent=15pt #2\par
        \parindent=15pt #3
        \end{minipage} }\par}}
 
\def\keywords#1{{
        \centering{\begin{minipage}{4.5in}\baselineskip=10pt\eightrm
        {\eightit Keywords}\/: #1
        \end{minipage} }\par }}
 
\renewenvironment{thebibliography}[1]                   
        {\ninerm
         \baselineskip=11pt                             
         \begin{list}{\arabic{enumi}.}
        {\usecounter{enumi}\setlength{\parsep}{0pt}
         \setlength{\leftmargin 17pt}{\rightmargin 0pt} 
         \setlength{\itemsep}{0pt} \settowidth          
        {\labelwidth}{#1.}\sloppy}}{\end{list}}
 
\newcounter{itemlistc}
\newcounter{romanlistc}
\newcounter{alphlistc}
\newcounter{arabiclistc}

\newcommand{\fcaption}[1]{
        \refstepcounter{figure}
        \setbox\@tempboxa = \hbox{\eightrm Fig.~\thefigure. #1}
        \ifdim \wd\@tempboxa > 5in
           {\begin{center}
        \parbox{5in}{\eightrm \smalllineskip Fig.~\thefigure. #1 }
            \end{center}}
        \else
             {\begin{center}
             {\eightrm Fig.~\thefigure. #1}
              \end{center}}
        \fi}
 
\newcommand{\tcaption}[1]{
        \refstepcounter{table}
        \setbox\@tempboxa = \hbox{\eightrm Table~\thetable. #1}
        \ifdim \wd\@tempboxa > 5in
           {\begin{center}
        \parbox{5in}{\eightrm\smalllineskip Table~\thetable. #1 }
            \end{center}}
        \else
             {\begin{center}
             {\eightrm Table~\thetable. #1}
              \end{center}}
        \fi}
 
\def\@citex[#1]#2{\if@filesw\immediate\write\@auxout    
        {\string\citation{#2}}\fi                       
\def\@citea{}\@cite{\@for\@citeb:=#2\do                 
        {\@citea\def\@citea{,}\@ifundefined             
        {b@\@citeb}{{\bf ?}\@warning
        {Citation `\@citeb' on page \thepage \space undefined}}
        {\csname b@\@citeb\endcsname}}}{#1}}
 
\newif\if@cghi
\def\cite{\@cghitrue\@ifnextchar [{\@tempswatrue
        \@citex}{\@tempswafalse\@citex[]}}
\def\citelow{\@cghifalse\@ifnextchar [{\@tempswatrue
        \@citex}{\@tempswafalse\@citex[]}}
\def\@cite#1#2{{$\null^{#1}$\if@tempswa\typeout
        {IJCGA warning: optional citation argument
        ignored: `#2'} \fi}}

\def\pmb#1{\setbox0=\hbox{#1}
        \kern-.025em\copy0\kern-\wd0
        \kern.05em\copy0\kern-\wd0
        \kern-.025em\raise.0433em\box0}

\def\fnt#1#2{\footnotetext{\kern-.3em
        {$^{\mbox{\scriptsize #1}}$}{#2}}}
 
\def\fpage#1{\begingroup
\voffset=.3in
\thispagestyle{empty}\begin{table}[b]\centerline{\footnotesize #1}
        \end{table}\endgroup}
 
\def\runninghead#1#2{\pagestyle{myheadings}
\markboth{{\eightit{\quad #1}}\hfill}{\hfill{\eightit{#2\quad}}}}
 
\font\tenrm=cmr10
\font\tenbf=cmbx10
\font\tenit=cmti10
\font\tenit=cmti10
\font\bfit=cmbxti10 at 10pt
 1
 1
 1

\font\ninerm=cmr9

\font\eightrm=cmr8
\font\eightit=cmti8

\def\qed{\hbox{${\vcenter{\vbox{                          
   \hrule height 0.4pt\hbox{\vrule width 0.4pt height 6pt
   \kern5pt\vrule width 0.4pt}\hrule height 0.4pt}}}$}}

\runninghead{M.Argollo de Menezes \& T.J.P. Penna}{Hopfield-Tank approach for the traveling salesman problem}

\begin{document}
\normalsize\textlineskip
{\thispagestyle{empty}
\setcounter{page}{1}
 
\renewcommand{\thefootnote}{\fnsymbol{footnote}} 
\def\bsc{{\sc a\kern-6.4pt\sc a\kern-6.4pt\sc a}}
\def\bflatex{\bf L\kern-.30em\raise.3ex\hbox{\bsc}\kern-.14em
T\kern-.1667em\lower.7ex\hbox{E}\kern-.125em X}
 
\copyrightheading{Vol. 0, No. 0 (1997) 000--000}
 
\vspace*{0.88truein}
 
\fpage{1}
\begin{center}{\bf IMPROVING THE HOPFIELD-TANK APPROACH}\\
{\bf FOR THE TRAVELING SALESMAN PROBLEM}
\end{center}
\vspace{0.37truein}
\centerline{\footnotesize M.Argollo de Menezes and T.J.P.Penna}
\vspace*{0.015truein}
\centerline{\footnotesize\it  Instituto de Fisica, Universidade Federal Fluminense}
\centerline{\footnotesize\it e-mail: marcio@if.uff.br and tjpp@if.uff.br}
\baselineskip=10pt
\centerline{\footnotesize\it  Av. Litor\^anea s/n,24210-340, Niter\'oi, RJ,
Brasil}
\vglue 10pt
\publisher{()}{()}
 
\vspace*{0.21truein}
\abstracts{\noindent
In this work we revisit the Hopfield-Tank algorithm for the traveling salesman problem 
(TSP)~\cite{hopfield} and report encouraging results, with a different dynamics, that
 makes the algorithm more efficient finding better solutions in much less computational
 time.}{}{}
 
\bigskip
\keywords{neural networks, optimization problems, traveling salesman problem, Kawasaki
 dynamics}
\vspace*{1pt}\textlineskip
\noindent
\textheight=7.8truein
\setcounter{footnote}{0}
\renewcommand{\thefootnote}{\alph{footnote}}
\par The traveling salesman problem (TSP) is the problem of finding the shortest path
 needed for a salesman to visit $N$ cities on a planar surface. It belongs to a class
 of problems called non-deterministic polynomial (NP) complete that have many
 applications to physics, mainly because of the methods developed to solve them. The
 computational complexity of the TSP problem resides in the rapidly increasing number 
of possible configurations with the number of cities. For example, for an N-cities
 problem, the number of possible configurations is $(N-1)!/2$, which is of order of
 $10^{1217}$ for $N=532$, therefore it is of vital importance to develop some heuristics
 to solve the problem and find near optimal solutions.
\par Among the methods employed to solve the TSP problem, there exists neural-networks
 (NN)~\cite{hopfield}~\cite{tjpp}, simulated annealing (SA) ~\cite{anneal}, genetic
 algorithms (GA) ~\cite{GA}, Lin-Kernighan elastic algorithm ~\cite{lin}, real-space
 renormalization ~\cite{japa} and others.
\par First proposed by Hopfield and Tank ~\cite{hopfield} in 1985, the neural-network
 realization of the TSP problem is of particular interest because of its ``silicon
 implementation''. However in other article, Wilson and Pawley ~\cite{cyber} discussed
 the stability of the method and reported unsatisfactory results even with some
 modifications on the original algorithm. In this paper we revisited this problem, and
 show that a Kawasaki dynamics of updating spin variables of the network is highly
 efficient in this problem. This article is divided in the following sections: in
 section two we summarize the Hopfield-Tank model. In section three the computational
 implementation of both the original algorithm and the new one are discussed and
 results are presented in section four.

\section{Hopfield neural-networks}
\par In a neural network we have the neurons that have a certain degree of activation,
 depending on how and with which intensity they are interconnected ~\cite{hopfield}.
 This connection between the neurons are called synapses, and they are not physical,
 but chemical. The degree of activation of a particular pair of neurons is proportional
 to the amount of neurotransmitter released by the neuron. This neurotransmitter allows
 the electric pulse generated by some stimulus to pass from one neuron to others,
 transmitting the information neuron by neuron until it reaches the brain. The neuron
 dynamics will be governed by  the state of other neurons which it is connected to
 by the synapses.
\par Among the many neural network models developed to simulate some features of human
 brain like recognition and learning, Hopfield ~\cite{hopf_net} proposed a neuron
 dynamics which minimize the following energy function
$$ H=-\sum d_{ij}\sigma_i \sigma_j. $$
where $\sigma_i$ is the state of the $i{th}$ neuron of the network and $d_{ij}$ is the
 synaptic strength between neurons $i$ and $j$. The dynamics which minimizes this
 energy function is 
$$ \sigma_i(t+1)={\rm sgn}(\sum_j d_{ij}\sigma_j)$$
 One first simplification from the behaviour of real neurons is to set the state of the
 neuron as a Boolean variable, that means, it is activated or not ($\sigma_i=-1,1$).
 One  particular feature of Hopfield-like neural networks is that they can modify their
 behaviour in response to their environment (feedback mechanism),that means, given a
 set of inputs to the neurons, they self-adjust to produce consistent responses. In
 order to simulate the features of the human brain, one has to implement a learning
 stage. One of the most used training algorithms is the Hebb rule, which is an 
unsupervised learning where the synaptic strength between two neurons is increased when
  both the source and destination neuron are activated. We can write this as
 $d_{ij}=\sum_{\mu}^N \xi_i^{\mu} \xi_j^{\mu}$ ,where $\xi^{\mu}$ is one of the $N$
 patterns stored in the brain. We represent one pattern by a set of $N$ neurons in a
 particular configuration.
\par This learning stage is appropriate for pattern recognition by a neural network, which
 means searching for a local minimum in the energy surface of $H$, which is different of the
 problem of optimization, where the search is for the global minimum. 
\par As this network device works as a multiple-feedback circuit, which is not governed by any
 variable parameter, like the artificial temperature in SA or mutations in GA, it is claimed as
 ``silicon implementable'', that is, it is possible to develop microchips that execute this
 function.
\par The realization of the traveling salesman problem in Hopfield neural-networks consists on
 attributing
 to the distance between two cities $A$ and $B$ a symmetric synaptic strength $d_{ab}$ and so on
 to all pairs
 of cities on a unit square (we normalize the maximum distance, and then the solutions given can be compared with any
 others with no need to rescaling). The problem is mapped onto a network by representing each city by a row of
 $N$ elements, where $N$ is the number of cities. Then, the output of a neuron, $\sigma_{ij}$ (called spin variable),
 means  the $i{th}$ city visited in the $j{th}$ order. If it is set to one, the city is really visited in that order, if it is set to zero, then this assertion is false.

\begin{figure}[htbp]
\epsfysize=2.0truein
\centerline{\epsfbox{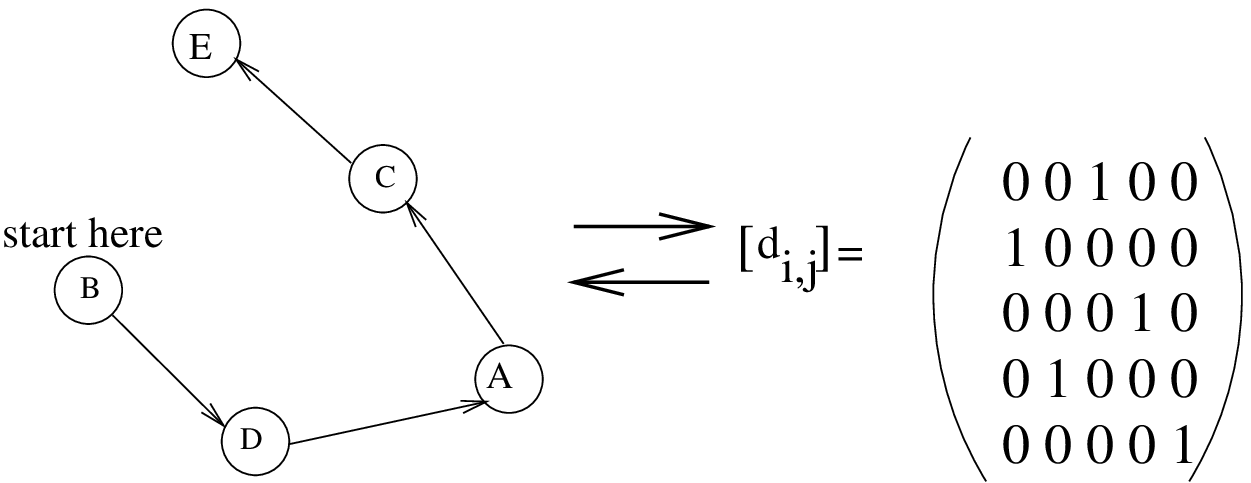}}
\fcaption{Matrix representation of the TSP problem}
\end{figure}

Fig. 1 is an example where the city $B$ is visited first, followed by $D$,$A$,$C$ and
 $E$. The resulting perimeter, or energy, is $d_{B,D}+d_{D,A}+d_{C,E}+d_{E,B}$. It can
 be verified that this $N\times N$ matrix is sparse (more exactly, it has only $N$
 elements).
\par The way of defining the energy function is not unique and requires a specification
 of the problem and what is to be minimized. In the TSP case, it is required that each
 row and each column have only one non-zero element (which means that a city is to be
 visited only once and only one city per time is visited), and the basic requirement is
 that solutions with short paths must be favoured. One satisfactory energy function
 proposed in ~\cite{hopfield} is
$$ E=\frac{A}{2}\sum_X^N\sum_i^N\sum_{j\neq i}^N \sigma_{Xi}\sigma_{Xj}$$ 
$$+ \frac{B}{2}\sum_i^N\sum_X^N\sum_{Y\neq X}^N \sigma_{Xi}\sigma_{Yi}$$  
$$+ \frac{C}{2}[(\sum_X^N\sum_i^N\sigma_{Xi})-N]^2 $$ 
$$+\frac{D}{2}\sum_X^N\sum_{Y\neq X}^N\sum_i^Nd_{XY}\sigma_{Xi}(\sigma_{y i+1}+\sigma_{y 
i-1})$$
with $A,B,C$ and $D$ $> 0$. The first and second terms vanish only when there is one and only
 one non-zero element in each row and column, respectively. The third term is zero when there
 are just $N$ of these non-null elements in the matrix, avoiding the convergence to the situation
 where all elements are zero, and the last term represent the length of any valid tour, and
 favours the shortest tours. The output of a neuron can be considered as a continuous variable,
 as in the original model, or discrete and equal to $0$ or $1$, as in this study, which is more
 efficient to deal with because of the well known computer tricks associated with these Boolean
 variables ~\cite{PMCO}.
\par In their 1985 paper ~\cite{hopfield}, Hopfield and Tank reported success for a 30 cities
 problem, a fact which is contested by Wilson and Pawley three years later $[1]$, who
 say that the convergence of the network to an acceptable tour, which means a tour with all
 constraints satisfied, is difficult to reach and that this possibility decreases rapidly with
 the number of cities (in fact, for a 10 cities problem, after 1000 iterations of the network an
 acceptable tour was obtained in only $8 \%$ of

 all cases). They also tried to improve the algorithm in three ways: by giving excessively large
 values to impossible spin configurations like $\sigma_{ii}$, varying the values of the energy
 parameters, and  changing the initial conditions. Even with all this changes the final result
 was not better than the original one, and since then interest on Hopfield NN algorithm for TSP
 has decreased. With the new trick shown in this paper we shed a new light on this problem.

\section{Computational Implementation}

\par In this section we will discuss the computational implementation of the problem, the
 properties of recurrent networks, like stability, and the improvement of the Hopfield-Tank 
algorithm for the traveling salesman problem by introducing the Kawasaki dynamics.
\par As cited on the previous section, Hopfield neural networks are said to be recurrent
, because of the dynamical feedback mechanism of changing the input as a function of the output
 of the network. For a stable network, after successive iterations the changes on the output
 will become smaller and smaller until it reaches an equilibrium state where all the outputs
 will become constant in the successive iterations. A network can be proven to be
 stable~\cite{book} if the synaptic matrix is symmetric with zero

s on its main diagonal, that means, $d_{ij}=d_{ji}$, and $d_{ii}=0$. But because of the highly
 irregular form of the TSP energy surface, there is no guarantee that it will converge to the
 global rather then to a local minimum, and even whether the minimum found will respect all the
 constraints (providing an acceptable tour) or not. The way to deal with it is to make a
 statistics of the $N$ cities problem.
\par With the constraints imposed, we can randomly change the state of a neuron and calculate
 the new energy, accepting the new configuration if it leads to a lower energy. In order to find
 minimal solutions we apply a parallel dynamics on the algorithm, that means, we change the
 state of a neuron and calculate the new energy, if it decreases we accept the change, if not,
 we invert it again. But the trick is to test all neurons at the same iteration, changing from
 the old configuration to the state obtained

 after all neuron have been tested and updated, and the results found are better then when
 updating sequentially: one single neuron at a time. We consider one time step when the
 configuration of the network doesn't change after $10\times N$ iterations, where $N$ is
 the number of cities, or if it does not converge after $100\times N$ iterations. 
This technique is slow, given the number of possible configurations of the spins, namely
 $2^{N^2}$, which increases enormously with the number of cities. Also, the number of operations
 required to determine the energy of a configuration is rather high.
\par What we can do to simplify this problem is try to reduce the number of operations of the
 algorithm. Our proposal is to update spin variables according to the Kawasaki dynamics. The
 Kawasaki dynamics consists in choosing two neurons and checking if their exchange will decrease
 the energy value. In the particular case of the TSP it can violate the constraints. Therefore,
 if we select only active neurons, we can start a simulation from a situation where there is
 only one active neuron in each column and row (like, for instance, the identity matrix), which
 means a valid tour. With this, we are able to always interchange two rows at a time, thus
 always having a valid tour as output and overcoming the problem of convergence. This task 
reduces the number of operations required to calculate the energy function, since we do not need 
the constraints anymore, and the computational time required to obtain a desired number of valid
 tours is greatly reduced.

\section{Numerical Results and Conclusions}
\par In order to have reliable results we make a statistics of the possible energy configurations
 (perimeter of the valid tours) of one city set, or make a statistics of the number of 
convergences to a valid tour for various randomly chosen city sets. As the number of acceptable
 tours increases with the number of cities, make a statistics of the possible energy 
configurations for a single city set is appropriate only when is $N$ large . A statistics of the
 number of convergences for a valid tour can give good results in comparing the efficiency of 
 different dynamics not for one single particular city set, but for as 

many sets as one can build. In this paper we will use the second statistics for 5000 different
 sets.
\par Simulating 5000 different city sets we can construct histograms for the final configurations
 obtained with both the traditional and Kawasaki dynamics and observe that this new task is able
 to find more near-optimal solutions and paths even shorter than the shortest path found in the
 traditional way, which explains the larger dispersion for the Kawasaki dynamics. It can be seen
 that it obeys a Gaussian distribution in both cases and that the number of possible states is
 greater for $N=20$.
\begin{figure}[htbp]
\epsfysize=2.7truein
\centerline{\epsfbox{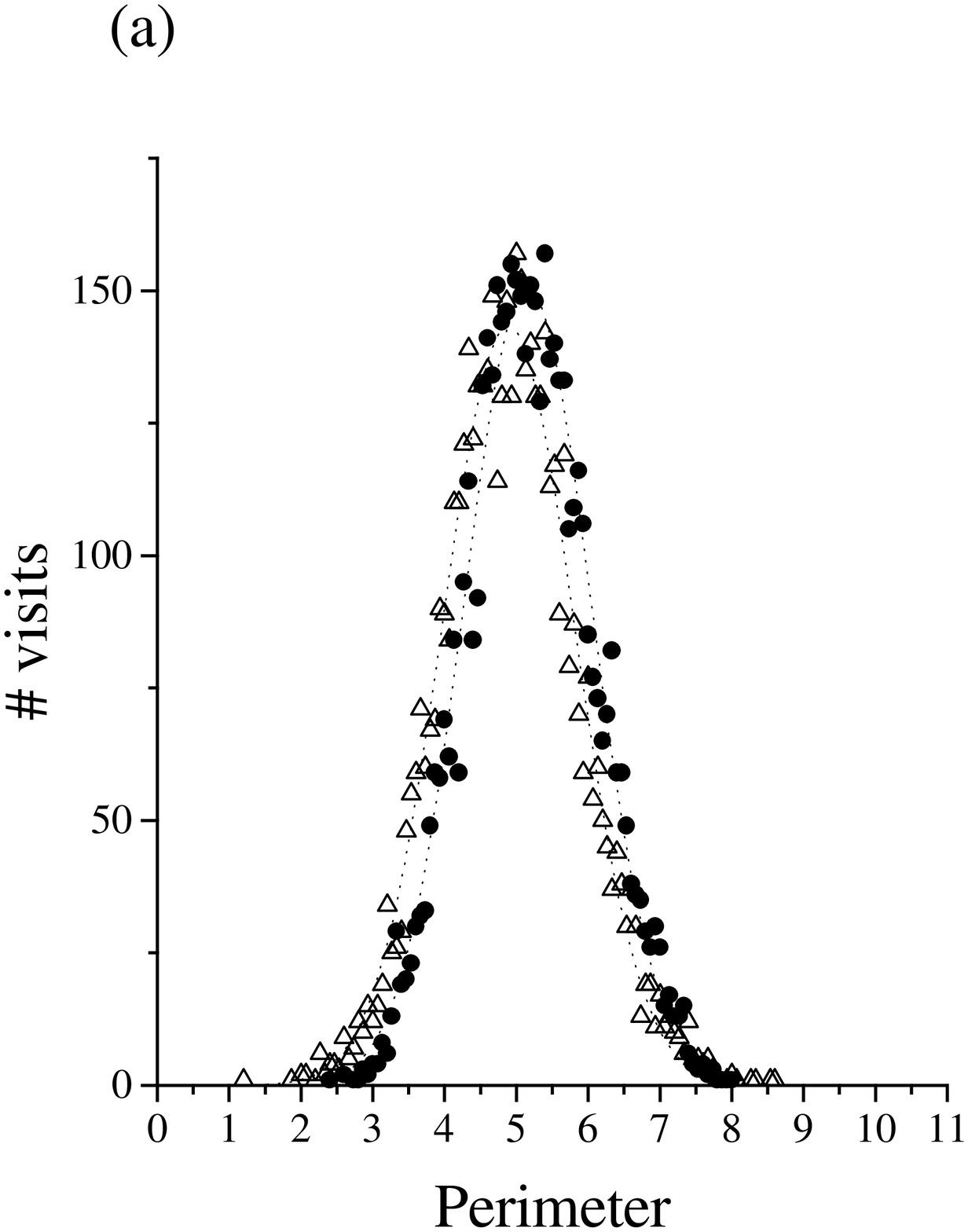}}
\epsfysize=2.5truein
\centerline{\epsfbox{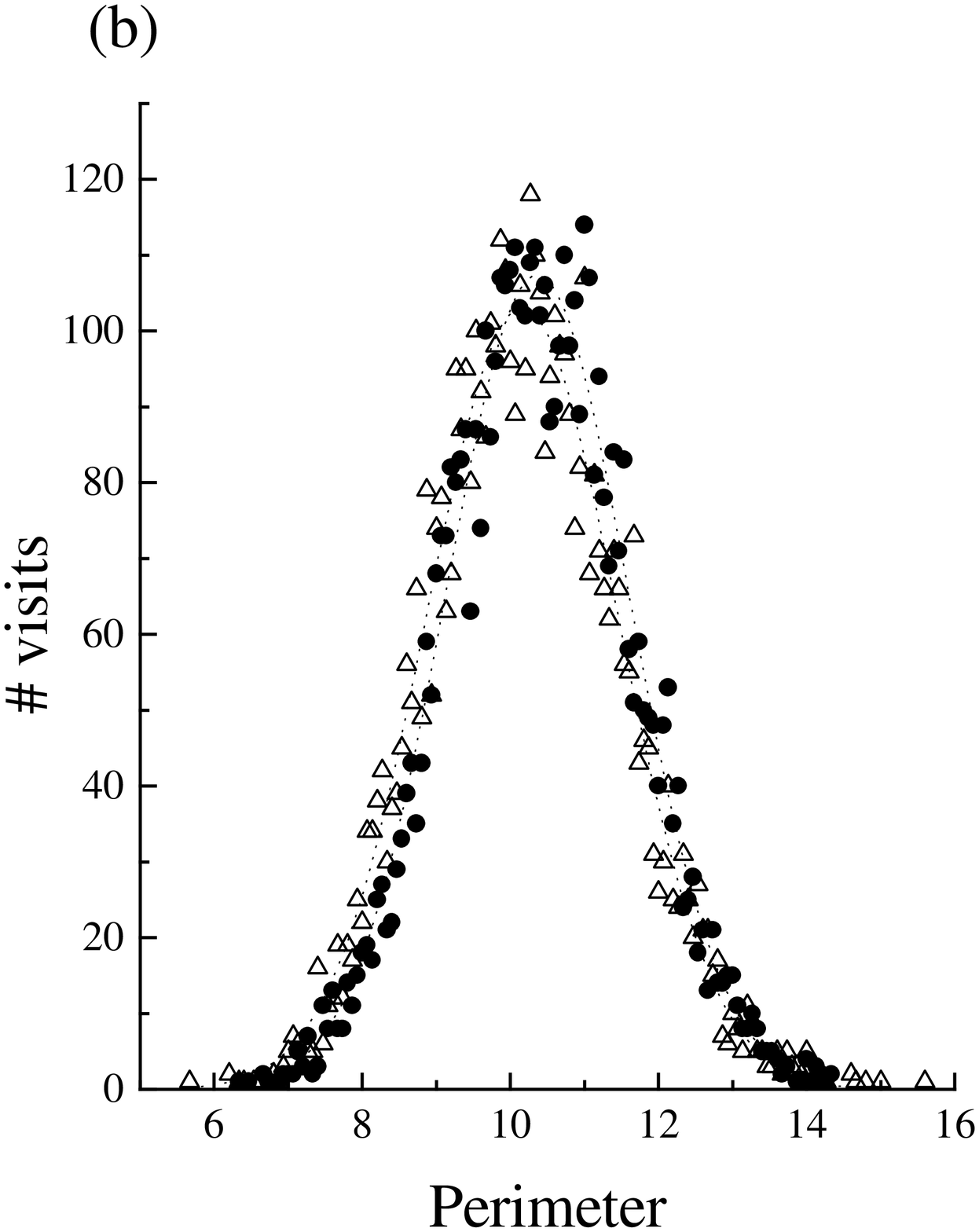}}
\fcaption{Histograms of number of convergences vs. perimeter ($P$) of the tour (energy) for
 {\bf (a)} 10 cities and {\bf (b)} 20 cities. Bullets are for Hopfield-Tank dynamics and
 triangles are for Kawasaki dynamics. Gaussian fits  for the histograms give ${\bar P_{HT}}=5.15$ and ${\bar
 \Delta P^2_{HT}}=1.77$ and ${\bar P_{K}}  =4.88$ and  ${\bar \Delta P^2_{K}}=1.85$ for $N=10$. For $N=20$
 we have ${\bar P_{HT}}=10.36$ and  ${\bar \Delta P^2_{HT}}=2.48$ and ${\bar P_{K}}  =10.15$ and  ${\bar
 \Delta P^2_{K}}=2.58$.}
\end{figure}

The other important feature of this method, the computational efficiency, is shown in fig.3, where the time is
 in seconds and was obtained in a Pentium 133 MHz, with the program in C code compiled with GNU C compiler, for
 200 sets . The results are better for $N$ large, which is the region of real interest. The fact that the
 difference between the histograms of the two dynamics is not so significant for $N=20$ is not disappointing,
 since the shortest path is always achieved with the Kawasaki dynamics and the time required for constructing
 the histograms are very different.
\begin{figure}[htbp]
\epsfysize=2.7truein
\centerline{\epsfbox{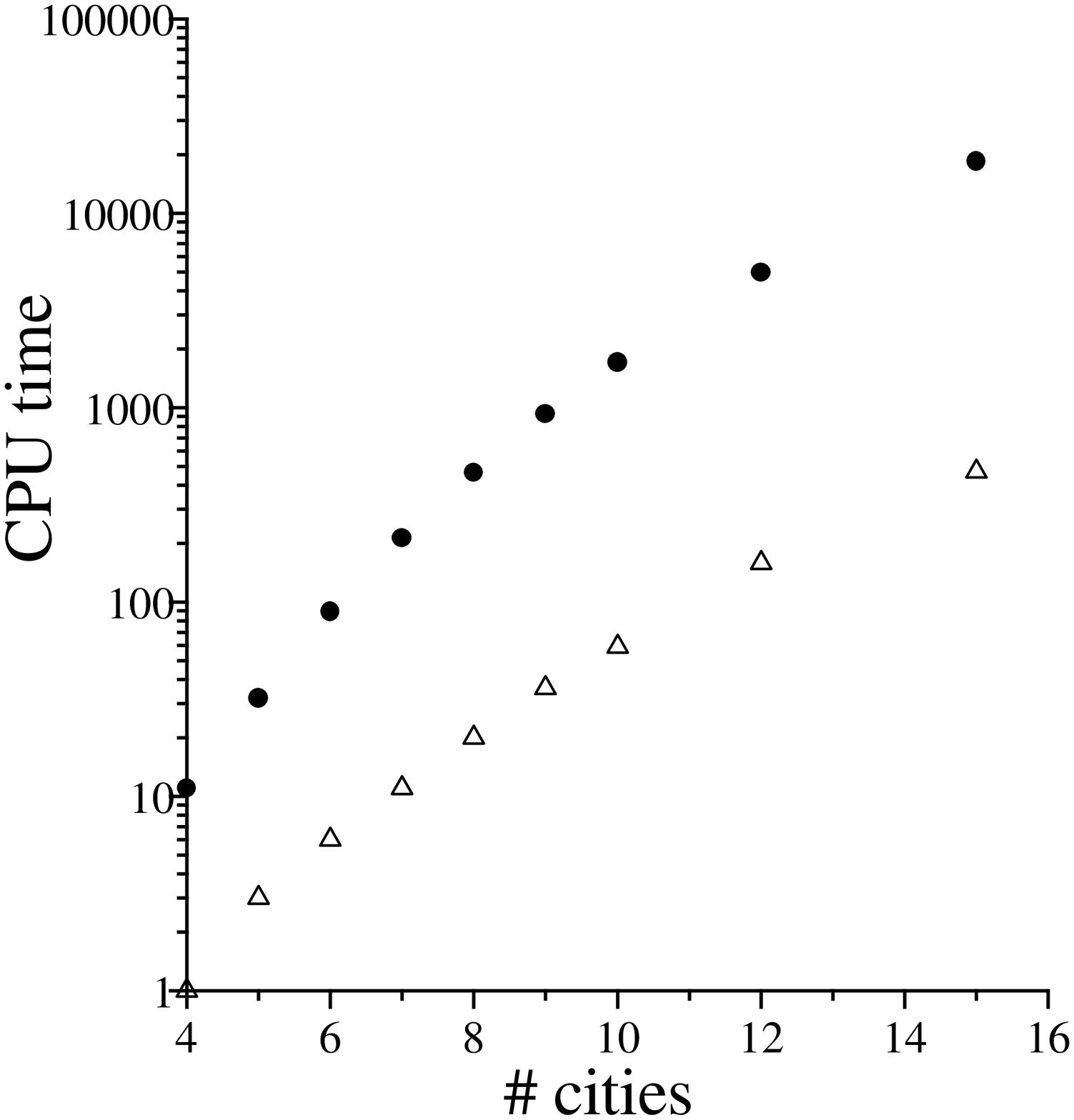}}
\fcaption{Computation time vs. number of cities for Hopfield-Tank ($\bullet$)  and Kawasaki
 dynamics ($\triangle$). Note the logarithm scale in the vertical axis.}
\end{figure}
 \par From the results shown above, it can be seen that this dynamics, using multi-spin coding
 techniques ~\cite{PMCO}, makes it worth applying the Hopfield NN algorithm to systems with 
sizes of real interest.

\section{Acknowledgements}
This work is partially supported by Brazilian agencies CNPq and CAPES.

\end{document}